# Observation of the First-Order Interference Fringes Beyond Coherence Length Employing Commercial Continuous-wave Multi-mode Laser Diode: A Sight of Two-photon Interference

Hongmin Liu

National Space Science Center, Chinese Academy of Sciences

No.1, ZhongguancunNanertiao, Haidian District. Beijing 100190, China

We report an experiment of observation of classical double-slit interference fringes of two-photon interference. In the experiment, a commercial continuous-wave multi-mode F-P laser diode without either mode-locked or frequency-locked is used as the light source, the density of photons is far more than the single-photon level, and the path difference of the long and short paths is far more than the longitudinal coherence length of the laser diode. The temporal stable and clearly visible spatial-distributed pattern, *i.e.* first-order interference fringes, was observed .Contradict to the prediction of single-photon interference mechanism, the interference happened far beyond the coherence length, the occurrence possibility and time duration of the interference fringes decrease with the reduction of mode number of the LD, and the time difference between the fringes disappearance and the light vanish is equal to the time difference of the two paths. After discussion, we came to the conclusion that the observed phenomena can be understood in time-resolved two-photon interference mechanism. We reveal a new method to perform two-photon first-order interference, and this help to understand the nature of two-photon interference and also can be useful for quantum information science.

Since Young's classic double-slit interference experiment, photon interference has been an important research in physics and has played an essential role in our understanding of the superposition principle and wave-particle duality in quantum mechanics, such as Dirac's statement on photon interference: "Each photon then interferes only with itself. Interference between different photons never occurs." [1], and Feynman's statement : "has in it the heart of quantum mechanics. In reality, it contains the only mystery."[2]. The mystery of photon interference inspire people to conduct photon interference experiments and observations with new strategies and apparatus continuously , both in classical and quantum mechanics.

Many experiments showed the single particle interference, such as photon, electron and proton[3][4] [5][6][7], meanwhile, there has been continuous research of two-photon interference both in experiment and theory. In 1963, Mandel *et al.* reported the observation of transient interference fringes with two independent optical master triggered simultaneously, and the measurement time was significantly shorter than the reciprocal spectral bandwidth (or coherence time) of the involved light sources[8]. In 1987, The most interesting and important two-photon interference experiment employing two identical photons from the spontaneous parametric down conversion (SPDC) process was reported by Hong, Ou, and Mandel(the HOM interference) [9],which is the clear evidence of two-photon interference. After that, many two-photon interference phenomena were founded, such as quantum beating[10][11] [12], induced interference[13] and subwavelength interference[14][15]. Now the conceptual and experimental frame work of two-photon interference is the basis of quantum information science/technology.

In classical physics, the coherence of light field is the spatial and temporal coherence of electromagnetic field, which is described in coherence length, coherence time, coherence area and volume. In quantum mechanics, the light field is the photon field, employing operators to quantize classical electromagnetic vector. In 1963, Glauber proposed the quantum correlation functions[16][17], which is used to describe the quantum interference. In single-photon interference, *i.e.* the first-order interference, both classical and quantum physics present the same calculation results, such as the expected fringe spacing in Young's double-slit interference experiment. In two-photon interference, *i.e.* the second-order interference, they present different calculation results, and the visibility of $V = 0.5$ in HOM interference is usually considered as the border between classical and quantum physics.

Most of the two-photon interference experiments utilizing single-photon counting and correlation techniques to observe the intensity correlation of two independent quantum light sources, which is either entangled photon pairs or significantly attenuated to the single-photon level. Recently research and studies show that classical light source can be used to perform two-photon interference experiments[18][19][20][21], and can be used to analog quantum interference in some conditions. While quantum calculation and information technology are making great progress both in research and application, and classical light sources are easier to operated and chipper than quantum sources, the research and experiments employing classical light source will be great help to the wide application of quantum calculation and quantum information technology.

In this paper, we report an experiment of observation of the classical double-slit interference beyond the coherence length employing commercial continuous-wave multi-mode F-P laser diode(LD) and optical fiber, the density of photons in the long and short paths is far more than the single-photon level. There are three parts in this experiment: part one is the observation of double-slit interference fringes, part two is the observation of the relationship between the visibility, *i.e.* the occurrence possibility and time duration, of interference fringes and the number of modes of the LD, and part three is the observation of the time relationship between the fringes disappearance and the light vanish on the screen. In the duration of observation which is far longer than the coherence time of the light source, the temporal stable and clearly visible double-slit fringes, *i.e.* first-order interference fringes, was observed. And the phenomena which is contradict to the physical mechanism of the single-photon interference were also recorded. After the discussion of the observed phenomena, we came to the conclusion that it is the results of time-resolved two-photon first-order interference.

Figure 1 shows the schematic setup to realize the experiment of double-slit interference observation. The light source is a commercial continuous-wave multi-mode F-P LD without either mode-locked or frequency-locked, $\lambda$=660nm and $P_o$=50mW; FBS is a 40:60 fiber beam splitter; $p_1$ is the short path(2m single-mode fiber HP-630); $p_2$, *i.e.* the fiber delay line, is the long path(600m and 1000m single-mode fiber G652D);FDS is a fiber double-slit(HP-630), 4μm slit width each and 125μm spacing between two slits; The distance between FDS and the screen is 31.5cm; $D_1$ and $D_2$ are SiPIN photon diode(S5973),sited at bright fringe and dark fringe separately; A&S is a circuit of amplifying and shaping，with Boolean "and" logic output "$D_1 \cdot D_2$" and Boolean "xor" logic output "$D_1 \oplus D_2$"; Timer is a time counter, counting "$D_1 \cdot D_2$" and "$D_1 \oplus D_2$" and generating the on/off signal to control the power switch K of LD( $t_{on}$ and $t_{off}$ are the moments at K is on and off separately); PSA is a power spectrum analyzer showing the power spectrum of the LD under a certain forward current.

Figure2 shows the sketch map of change of output of $D_1$ and $D_2$ over time, and the pattern variation on the screen alone with the time variation if the interference happened, *i.e.* the time relationship between the fringes disappearance and the light vanish on the screen. In figure 2, LD is lightless during the K is off, so the pattern on the screen is dark, the output of $D_1$ and $D_2$ is zero, "$D_1 \cdot D_2$" =0 and "$D_1 \oplus D_2$" =0. At the moment $t_{on}$ K is on, LD gives out light $\psi(t_{on})$; $\psi(t_{on})$ arrives at FBS after $t_0$ and is divided into two beams $\psi_1(t_0)$ and $\psi_2(t_0)$; passing through $p_1$, beam $\psi_1(t_0)$ arrives at FDS after $t_1$ and varies to beam $\psi_1(t_1)$, but beam $\psi_2(t_0)$ dose not arrive at FDS , there is only one beam, so the screen is bright, the output of $D_1$ and D2 is not zero, "$D_1 \cdot D_2$" =1 and "$D_1 \oplus D_2$" = 1; passing through $p_2$, beam $\psi_2(t_0)$ arrives at FDS after $t_2$ and varies to beam $\psi_2(t_2)$, so there two beams arrive on the screen, and if the interference happened there are interference fringes on the screen, the output of $D_1$ is not zero, the output of $D_2$ is zero, "$D_1 \cdot D_2$"=0, "$D_1 \oplus D_2$" = 1. At the moment $t_{off}$ K is off, LD gives out no light, $\psi(t_{off})$=0, and $\psi_1(t_0)$ =0 and $\psi_2(t_0)$=0 after $t_0$; $\psi_1(t_1)$=0 after $t_1$, but beam $\psi_2(t_1) \neq 0$ , there is only one beam, so the screen is bright, the output of $D_1$ and D2 is not zero, "$D_1 \cdot D_2$"=1, "$D_1 \oplus D_2$" =1; $\psi_2(t_2)$=0 after $t_2$, so there no beams arrive on the screen, the output of $D_1$ and $D_2$ is zero, "$D_1 \cdot D_2$"=0 , "$D_1 \oplus D_2$" =0.

Obviously, the time duration of "$D_1 \cdot D_2$"=1 is the time difference($\Delta t$) of the long path $p_2$ and the short path $p_1$, i.e. $\Delta t = n \cdot (p_2 - p_1)/c$, $n$ is the refraction index of fiber, $c$ is the light speed in vacuum; the time duration of "$D_1 \oplus D_2$"=1 is the time that interference happens. In our experiment, the time that light pass from LD to FBS is much less than it passing through $p_2$ and $p_2 \gg p_1$, so $\Delta t \approx t_2 - t_{on}$ or $\Delta t \approx t_2 - t_{off}$.

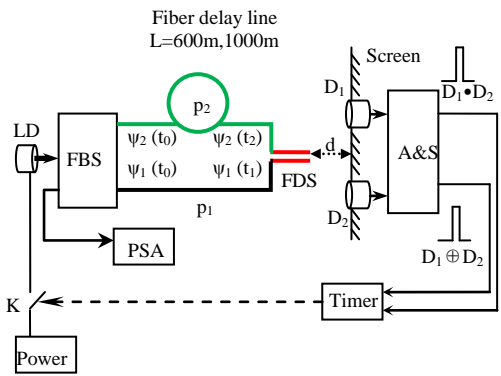

Fig. 1.schematic setup of the experiment.
LD: Laser Diode; K:power switch of LD; FBS:40:60 fiber beam splitter; FDS: fiber double-slit; d: distance between FDS and screen; $D_1$ and $D_2$: SiPIN photon diode; A&S: a circuit of amplifying and shaping; "•": Boolean "and" logic; "⊕": Boolean "xor" logic; timer: time counter; PSA: power spectrum analyzer.

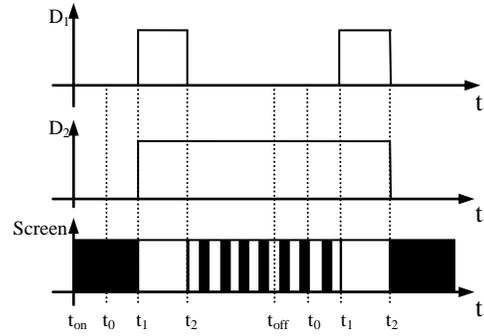

Fig.2.Sketch map of variation of the output of SiPINs and the pattern on the screen alone with the time variation.
$D_1$ and $D_2$:output of SiPINs, $t_{on}$: the moment K is on, $t_{off}$: the moment K is off, $t_0$: time of light passing from LD to FBS, $t_1$: the time beam $\psi_1$ passing through $p_1$, $t_2$: the time beam $\psi_2$ passing through $p_2$.

After the observation of fringes with $p_2-p_1$=0, the measurement of the power spectra of LDs and the time difference of the two pass of the circuit, the observation of the double-slit interference beyond coherence length was conducted.

In part one, the measurement time about 2 minutes and the temporal stable and clearly visible double-slit fringe patterns was obtained, shown in figure 3(a), and the interference fringes recorded with path difference is zero is given in figure 3(b). Obviously, the interference fringes shown in figure 3 is spatial-distributed pattern as the result of phase coherence of the light field, i.e. first-order interference. So we can calculate the spacing of the fringes with first-order interference theory and compare it to the recorded spacing. In figure 3, the measured fringes spacing is 1.67mm±0.05mm,while the corresponding value calculated from $x=\lambda d/w$, where $x$ is the expected spacing of fringes, $\lambda$ is the wave-length, $d$ is the slit-to-screen distance and $w$ is the slit spacing, we get $x$=1.663mm with $\lambda$=660nm, $d$=31.5cm, and $w$=125μm. The measured and calculated fringes spacing is equal in within the error permission.

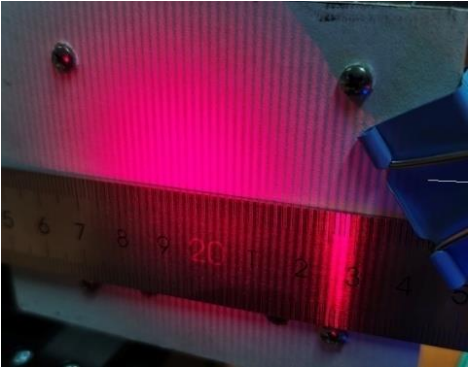

(a) First-order fringes recorded with $p_2-p_1$=600m,1000m.

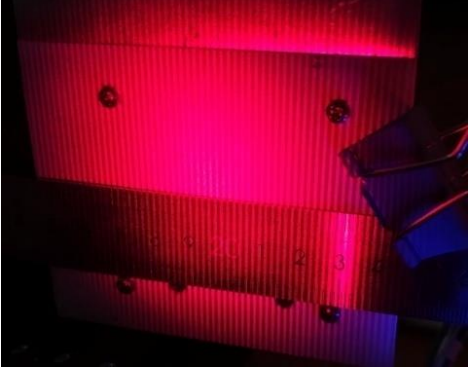

(b) First-order fringes recorded with $p_2-p_1$=0

Fig.3. An example of fringes recorded, The measured fringe spacing is 1.67mm±0.05mm.(a).$p_2-p_1$=1000m.(b).$p_2-p_1$=0.

According to the Wiener-Khintechine theorem, the coherence function of a light field $\gamma(\tau)$ is the inverse Fourier transform of the power spectra P(ν)[22],i.e. F($\gamma(\tau)$)=|E(ν)|$^2$=P(ν), so one can calculate the coherence length of LD from the measured power spectra of LD. In the one mode condition, the power spectra of LD can be written in the Gauss formation[23]:

$$p = p_0 \exp\{-[2\frac{(\ln 2)^{1/2}(\nu-\nu_0)}{\delta\nu}]^2\} \quad (1)$$

Where $\delta\nu$ is the line width of the half power point. With the calculation of inverse Fourier transform of (1), we have the modulus of the coherence function of a single-mode LD, i.e. the visibility of interference fringes:

$$V(\tau) = \exp\left[-\frac{(\pi\tau\delta\nu)^2}{4\ln 2}\right] \quad (2)$$

Where $\tau$ is the time difference of the two paths and the corresponding length difference of the two paths is $\Delta l = c\tau$, c is light speed。 According to the definition of coherence length, form(2),then the coherence length $l_c$:

$$l_c = \frac{0.624c}{\delta\nu} = \frac{0.624\lambda^2}{\delta\lambda} \quad (3)$$

Figure 4 shows the relation between visibility V and path difference $\Delta l$.

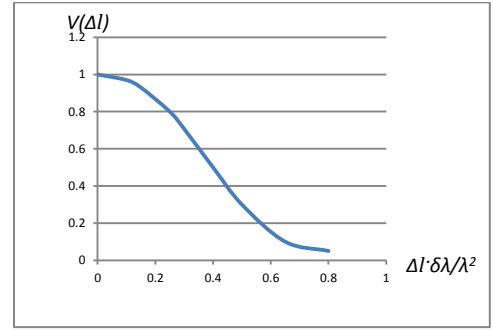

Fig.4. Visibility-$\Delta l$ curve of a single-mode light source

In the multi-mode condition, the frequency difference of two adjacent modes $\Delta\nu_q$ is $\Delta\nu_q=c/2L$, L is cave length of the LD, then the frequency difference of two modes is $k\Delta\nu_q$, k is positive integer for a LD. For simplicity, given the output power amplitude of each mode is the same $p_0$ and for the mode number is N, then the power spectra of a multi-mode LD can be written as:

$$p_N(\nu) = p_0 \sum_{n=-(N-1)/2}^{(N-1)/2} \delta[\nu - (\nu_0 + n + k\Delta\nu_q)] \quad (4)$$

Where $\nu_0$ is the centre frequency of the LD. For N is odd number, $n=\pm 1,\pm 2,\ldots,\pm(N-1)/2$; for N is even number, $n=\pm 1/2,\pm 3/2,\ldots,\pm(N-1)/2$. From (4), we have the modulus of the coherence function of a multi-mode LD:

$$|\gamma_N(\Delta l)| = \left|\frac{\sin(kN\pi\Delta l/2L)}{N\sin(k\pi\Delta l/2L)}\right| \quad (5)$$

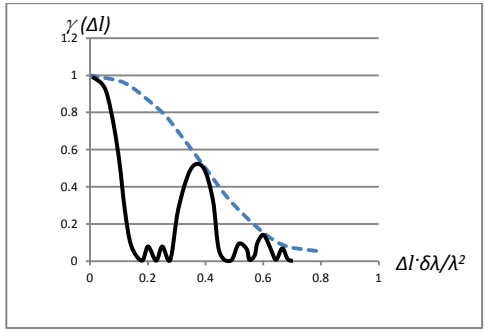

(a) relationship between $\gamma(\Delta l)$ and $\Delta l$

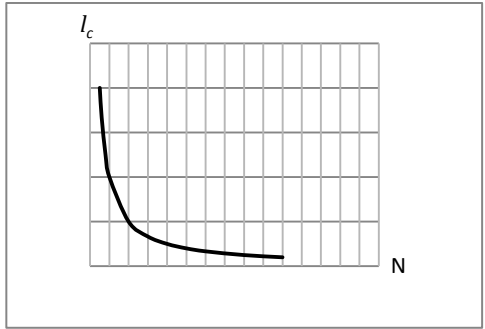

(b) relationship between $l_c$ and $N$

Fig.5. $\gamma(\Delta l)$-$\Delta l$ curve and $l_c$-$N$ curve of a multi-mode light source。

Figure 5 shows the schematic map of (5). In figure5 (a), the modulus of the coherence function of a multi-mode LD is a comb function over $\Delta l$, within the enveloping line which is the modulus of the coherence function of a single-mode LD. In figure5 (b), the coherence length $l_c$ decreases with the increase of $N$, i.e. the $l_c$ of a single-mode LD is longer than the one of a multi-mode LD. Although the results we get here, such figure5, is base on the assumption that the amplitude of each mode is the same, these can also be applied on the condition that the amplitude of each mode is different base on the same physics mechanism.

Figure 6 shows the measured output power spectra of the multi-mode F-P LD. In fig.6, the line width of the envelope the power spectra is about 1.5nm, and the line width of one mode is about 0.23nm, so the calculated coherence length according to (3) is about 0.29mm and 1.88mm separately. Then we have the conclusion that the $p_2$-$p_1$ (600m, 1000m)is far more than the coherence length of the light source when the first-order inference happens, which is contradict to the physical mechanism of the one-photon interference.

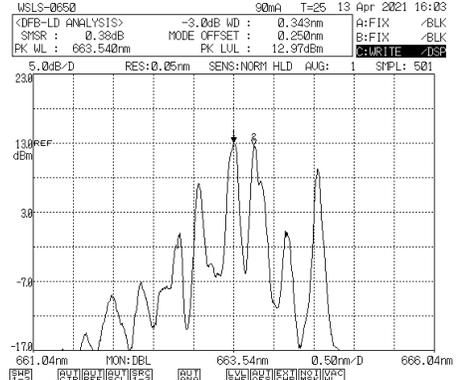

Fig.6. Measured output power spectra of the multi-mode F-P LD

In part two, the observation of the variation of occurrence possibility and time duration, i.e. the visibility, of fringes patterns with the mode number of LDs is performed. The number of modes can be changed easily by rising or lowing the forward current for a multi-mode F-P LD and by using a DFB LD. Figure 7 shows the decrease of mode number with lowing the forward current and a DFB LD .The power spectra in fig.7(a) belongs to the same LD in fig.6(a) but with the forward current at 50mA and the power spectra in fig.7(b) belongs to a DFB LD with the forward current at 100mA and the output power at 15mW.Obviously, the number of modes in the power spectra shown in figure 7 is less than it in figure 6. In the condition given in figure 7, the recorded phenomena is that the occurrence possibility and duration of interference fringes decrease with the reduction of mode number of the LD with $p_2$-$p_1$ is far more than the coherence length(600m, 1000m), but the interference fringes appears stably and clearly no matter how the number of modes changed with $p_2$-$p_1$=0. That the visibility of interference fringes decrease with the reduction of mode number of the LD is not agreement with (5),which is based on the physical mechanism of the one-photon interference.

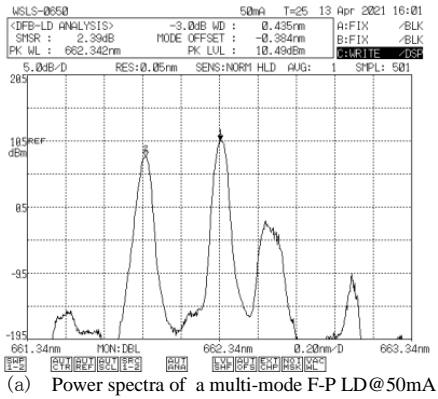
(a) Power spectra of a multi-mode F-P LD@50mA

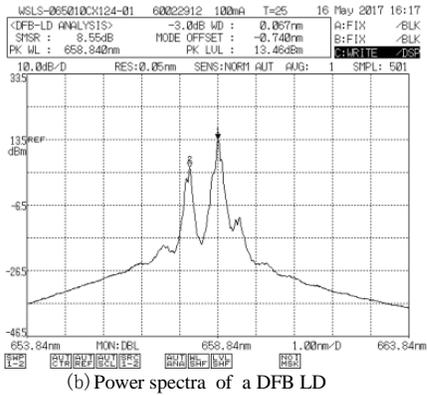
(b) Power spectra of a DFB LD

Figure 7. Power spectra of LDs with decreased number of modes.

In part three, the measured $\Delta t=t_2-t_{off}$ is 28-34μs or 48-52μs, corresponding to the value calculated from $\Delta t=n\cdot(p_2-p_1)/c$=29.3μs or 48.9μs separately, with $n$=1.4677 and $p_2-p_1$=600m or 1000m. The results of the measured $\Delta t$ and calculated $\Delta t$ show good agreement. The phenomenon that the time difference between the fringes disappearance and the light vanish on the screen was equal to the time difference of the two paths, shows the overlapped two light beams is emitted in different transition of the light source, which is also contradict to the conclusion that interference only happen with photon emitted in one transition of a light source according to the physical mechanism of the one-photon interference.

As have discussed above, we observed the first-order interference which is contradict to the physical mechanism of the first-order interference, but for two-photon interference there is no restriction of coherence length, coherence time, same one transition and so on, and two photons are emitted from different modes. So we can conclude that the observed first-order interference can be understood in the two-photon interference mechanism.

In summary, we reported the experiment of two-photon first-order interference with multi-mode LD. The temporal stable and clearly visible spatial-distributed pattern as the result of phase coherence of the light field, *i.e.* first-order interference is observed. Although the two-photon is time-resolved, but is not path-resolved after passing through FBS. In the experiment, a commercial F-P LD without either mode-locked or frequency-locked is used as the light source, so we reveal a new arrangement to perform two-photon first-order interference, and this help to understand the nature of two-photon interference and also can be useful for quantum information science.